\documentclass[aps,pra,twocolumn,superscriptaddress,floatfix,
nofootinbib,showpacs,longbibliography]{revtex4-1}
\usepackage[utf8]{inputenc}  
\usepackage[T1]{fontenc}     
\usepackage[british]{babel}  
\usepackage[sc,osf]{mathpazo}
\usepackage[scaled=0.86]{berasans}  
\usepackage[colorlinks=true, citecolor=blue, urlcolor=blue, anchorcolor=blue, linkcolor=blue ]{hyperref}  
\usepackage{graphicx}
\usepackage[babel]{microtype}  
\usepackage{amsmath,amssymb,amsthm,bm,amsfonts,mathrsfs,bbm} 

\usepackage{xspace}  
\usepackage{pgf,tikz}
\usepackage{xcolor}
\usepackage{multirow}
\usepackage{array}
\usepackage{bigstrut}
\usepackage{braket}
\usepackage{color}
\usepackage{natbib}
\usepackage{multirow}
\usepackage{mathtools}
\usepackage{float}
\usepackage{xcolor,colortbl}
\usepackage{physics}
\usepackage{amsmath}
\usepackage{color}
\usepackage[justification=justified, format=plain]{subcaption}
\usepackage[justification=raggedright]{caption}

\newcommand{\be}{\begin{equation}}
\newcommand{\ee}{\end{equation}}
\newcommand{\ba}{\begin{eqnarray}}
\newcommand{\ea}{\end{eqnarray}}

\newtheorem{theorem}{Theorem}

\newtheorem{definition}{Definition}

\newtheorem{observation}{Observation}
\newtheorem{example}{Example}

\newtheorem{lemma}{Lemma}

\usepackage{relsize}

\def\>{\rangle}
\def\<{\langle}

\begin{document}
	
\title{Detection of non-absolute separability in quantum states and channels through moments}	

\author{Bivas Mallick }
\email{bivasqic@gmail.com}
\affiliation{S. N. Bose National Centre for Basic Sciences, Block JD, Sector III, Salt Lake, Kolkata 700 106, India}

\author{Saheli Mukherjee}
\email{mukherjeesaheli95@gmail.com}
\affiliation{S. N. Bose National Centre for Basic Sciences, Block JD, Sector III, Salt Lake, Kolkata 700 106, India}

\author{Nirman Ganguly}
\email{nirmanganguly@hyderabad.bits-pilani.ac.in,nirmanganguly@gmail.com}
\affiliation{Department of Mathematics, Birla Institute of Technology and Science, Pilani, Hyderabad Campus, Jawahar Nagar, Kapra Mandal, Medchal District, Telangana 500078, India}

\author{A. S. Majumdar}
\email{archan@bose.res.in}
\affiliation{S. N. Bose National Centre for Basic Sciences, Block JD, Sector III, Salt Lake, Kolkata 700 106, India}

\begin{abstract}
In quantum information and computation, the generation of entanglement through unitary gates remains a significant and active area of research. However, there are states termed as absolutely separable, from which entanglement cannot be created through any non-local unitary action. Thus, from a resource-theoretic perspective, non-absolutely separable states are useful as they can be turned into entangled states using some appropriate unitary gates. In this work, we propose an efficient method to detect non-absolutely separable states. Our approach relies on evaluating moments that can bypass the need for full state tomography, thereby enhancing its practical applicability. We then present several examples in support of our detection scheme. We also address a closely related problem concerning states whose partial transpose remains positive under any arbitrary non-local unitary action. Furthermore, we examine the effectiveness of our moment-based approach in the detection of quantum channels that are not absolutely separating, which entails the detection of resource preserving channels. Finally, we demonstrate the operational significance of non-absolutely separable states by proving that every such state can provide an advantage in a quantum-channel discrimination task.

\end{abstract}
\maketitle

\section{Introduction}
The advent of quantum technologies has brought the use of quantum correlations to the forefront of modern-day research. Entanglement \cite{horodecki2009quantum} is one such quantum correlation underpinning communication advantages \cite{PhysRevLett.69.2881,PhysRevLett.70.1895,PhysRevLett.126.250501,piveteau2022entanglement,PhysRevLett.134.020802}, secure cryptography protocols \cite{PhysRevLett.67.661,lo1999unconditional,yin2020entanglement}, computational speed-up \cite{jozsa2003role,PhysRevA.101.012349,PhysRevLett.133.230604}, randomness generation \cite{pironio2010random,PhysRevLett.120.010503,PhysRevApplied.17.034011}, etc. This motivated the study of quantum entanglement from a resource-theory perspective, where the free states are the separable states ($\mathbb{S}$), the entangled states are the resourceful states, and the free operations are local operations assisted by classical communication (LOCC) \cite{chitambar2019quantum}. So, in the context of entanglement theory, the separable states do not offer any advantage in terms of their operational utility. However, certain separable states can be transformed into an entangled state by the choice of a suitable global unitary operation, and that entanglement can be harnessed to perform several information processing tasks. Mathematically, a global unitary operation results in a change in basis. A quantum state may be separable in one basis and entangled in another. 
However, there exist states from which no entanglement can be created by any global unitary operation. This constitutes the set of \textit{absolutely separable} states ($\mathbb{AS}$). These are the states that remain separable in any basis.

The concept of absolute separability for two qubits was introduced in \cite{PhysRevA.63.032307}. This was followed by characterising the global unitary operations that can maximise the entanglement content of two-qubit mixed states \cite{PhysRevA.64.012316}. Since the creation of entanglement lies at the cornerstone of many experimental protocols \cite{sackett2000experimental,PhysRevA.63.062309,lanyon2009experimentally,PhysRevLett.106.090502,yanamandra2025breaking}, the identification of non-absolutely separable states is of utmost importance. A necessary and sufficient detection criterion of non-absolute separability is based on the eigenvalue spectrum for qubit-qudit states \cite{PhysRevA.88.062330}. But, in order to implement this criterion, one needs to reconstruct the state by tomography \cite{RevModPhys.29.74,PhysRevLett.109.120403} to gain knowledge about the corresponding eigenvalues. From a geometric perspective, there exists a ball centered at the maximally mixed state such that all states lying inside the ball are absolutely separable \cite{PhysRevA.58.883,adhikari2021constructing}. However, there exist absolutely separable states outside the ball \cite{PhysRevA.62.022310}. For a state $ \rho \in \mathcal{D}(\mathbb{C}^d \otimes \mathbb{C}^d)$, the largest ball (maximal ball) of absolutely separable states is described by $\Tr(\rho^2) \le \frac{1}{d^2-1}$ \cite{PhysRevA.66.062311}. 

A  method of detecting absolutely separable states is based on characterizing the boundary of such states \cite{PhysRevA.103.052431,lewenstein2022linear,abellanet2025sufficient}. Other detection schemes involve the use of non-absolutely separable witnesses \cite{PhysRevA.89.052304,patra2021efficient,serrano2024absolute}. The set of absolutely separable states forms a convex, compact subset of the set of separable states \cite{PhysRevA.89.052304}. So, by the Hahn-Banach hyperplane separation theorem \cite{holmes2012geometric}, there exists a hyperplane that separates at least one non-absolute separable state from the set of absolutely separable states. Such witnesses can be constructed from entanglement witnesses by following the prescription in \cite{PhysRevA.89.052304,patra2021efficient}. While witness-based techniques can be applied to states of arbitrary dimensions, they generally require partial prior knowledge of the state, such as its entanglement class or an approximate parametric form, to design a suitable witness \cite{matsunaga2025detecting,guhne2009entanglement,guhne2006nonlinear}.

 In this work, we aim to identify the signature of non-absolute separability without requiring prior structural assumptions about the state. To achieve this, we introduce a moment-based approach that effectively bypasses the need for full state tomography. Although tomography does not require prior knowledge of the state, it typically demands an exponentially large number of measurement settings, making it experimentally costly. In contrast, our approach involving the evaluation of simple functionals can be efficiently estimated in experimental settings using the shadow tomography technique \cite{aaronson2018shadow,aaronson2019gentle,huang2020predicting,elben2020mixed}. Moreover, a polynomial number of state copies is sufficient for our purpose, whereas full state tomography-based techniques require an exponential number of copies. This highlights the significant advantage of our approach in terms of scalability as the system size increases. 
 
\begin{figure}[ht]
\includegraphics[width=.5\textwidth]{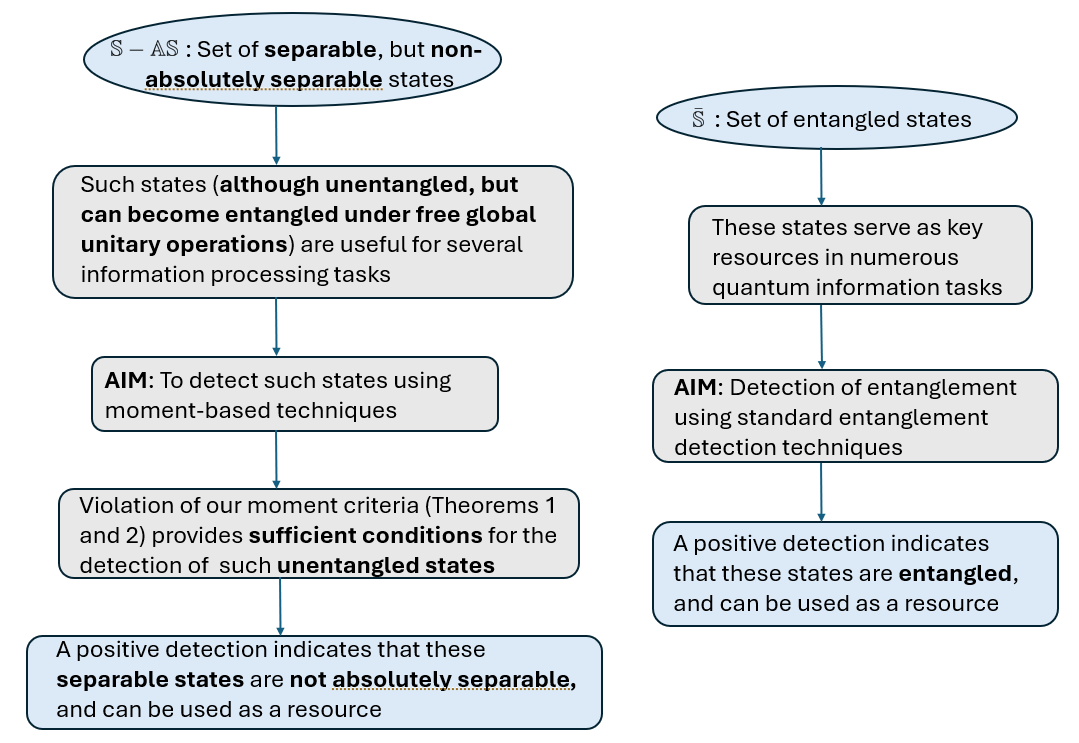} 
\caption{A flow diagram illustrating the basic distinction between the problem of entanglement detection, and the detection of non-absolutely separable states using our moment-based technique.}\label{fig1}
\centering
\end{figure}

 It is worth noting that the use of moments for entanglement detection, along with their experimental implementations, has been extensively studied in the literature \cite{gray2018machine,elben2020mixed,Neven2021,yu2021optimal,aggarwal2024theoretical,mukherjee2025detecting}. However, in all these works, the moments are employed to identify the entanglement present in a given initial quantum state. In contrast, our approach begins with a separable state and investigates the potential for entanglement generation from that state. A violation of our moment-based criterion thus identifies separable states that has the capability to generate entanglement under free global unitary operations. For better clarity, the basic distinction between entanglement detection and the detection of non-absolutely separable states using our approach is illustrated in Fig.\ref{fig1}. 
 
 A closely related problem is the detection of states with the absolute-PPT property, where PPT stands for positive under partial transpose. These are states whose partial transpose remains positive under any non-local unitary action. A necessary and sufficient condition (in terms of eigenvalues) for their detection was provided in \cite{PhysRevA.76.052325}. The conditions derived in \cite{PhysRevA.76.052325} hold for any dimension in the bipartite scenario. They are in the form of matrix inequalities; however, the size of the matrices increases tremendously with the increase in dimension. To overcome this issue, in this paper, we also detect non-absolute PPT states through the moment-based approach. We then present several illustrative examples to support our detection technique.

 We next examine the significance of non-absolutely separable states within the context of quantum channels. In certain situations, environmental noise can be so pronounced that it significantly diminishes the utility of a quantum state as a resource \cite{heinosaari2015incompatibility,srinidhi2024quantum,mallick2024characterization,kumar2025fidelity}. Some quantum channels possess the property that, regardless of the input state, the output is always absolutely separable. Such channels are known as absolutely separating channels \cite{filippov2017absolutely}. Therefore, from a resource-theoretic viewpoint, it becomes crucial to identify channels that are not absolutely separating, as these allow the possibility of recovering entanglement through an appropriate global unitary transformation. Motivated by this, in this present work, we develop a moment-based approach to detect such non-absolutely separating channels, thereby enabling the identification of channels that remain operationally useful for a broad class of quantum information protocols.
 
 Further, we explore the operational relevance of non-absolutely separable states in the context of quantum channel discrimination tasks. In a channel discrimination scenario, a quantum channel is randomly selected from a known pair with equal prior probability, and the objective is to determine which channel was applied, while minimizing the probability of error. It is well established that bipartite probe-ancilla quantum states can enhance the success probability in such tasks \cite{kitaev1997quantum}. Therefore, the most general strategy involves preparing a bipartite probe-ancilla quantum state, applying the given channel to the probe subsystem, and performing a two-outcome positive operator valued measurement (POVM) on the resulting joint state to infer which channel was used. Here, we ask the question: Can non-absolutely separable states offer an advantage in quantum channel discrimination tasks? We answer this in the affirmative by proving that every bipartite non-absolutely separable state can outperform all absolutely separable states in discriminating at least one pair of quantum channels. Specifically, for each non-absolutely separable state, there exists a channel discrimination scenario in which it achieves a strictly higher success probability than any absolutely separable state.

The structure of the paper is organized as follows: In Section \ref{s2}, we provide a concise overview of the preliminary concepts, including absolutely separable states, absolutely separable channels, and the moment-based criteria. Section \ref{s3} introduces our proposed framework for detecting non-absolutely separable states and non-absolute PPT states. Section \ref{s4} is devoted towards the detection of non-absolutely separating channels. In Section \ref{s5}, we investigate the operational advantage of non-absolutely separable states in the context of quantum channel discrimination. Finally, Section \ref{s6} concludes with a summary of our key results and discusses potential avenues for future research.

\section{Preliminaries}\label{s2}
Here, we introduce the basic notations used throughout the paper. Let $\mathbb{C}^d$ represent the finite, $d$-dimensional complex Hilbert space. The set of linear operators acting on a $d$-dimensional complex Hilbert space is represented by $d \cross d$ matrices, denoted by $\mathcal{M}_d$. Linear, positive maps acting on  $d \cross d$ matrices are denoted by $\Lambda : \mathcal{M}_d \rightarrow \mathcal{M}_d$. Quantum states, namely density operators, are a set of positive operators with unit trace acting on $\mathbb{C}^d$. This set is represented by $\mathcal{D}(\mathbb{C}^d)$. For composite (bipartite) systems, this state space consists of separable and entangled states. Let $\mathbb{S} \subset \mathcal{D}(\mathbb{C}^d \otimes \mathbb{C}^d)$ and $\mathbb{\bar{S}} \subset \mathcal{D}(\mathbb{C}^d \otimes \mathbb{C}^d)$ denote the set of bipartite separable and entangled states, respectively. 
\subsection{Absolutely separable states}
Consider the set of all bipartite separable states. Within this set, \textit{absolutely separable states} form a subclass of separable states that remain separable under the action of any global unitary, i.e.,
\begin{equation}
    \mathbb{AS} = \{\rho : U \rho U^{\dagger} \in \mathbb{S} \hspace{2mm} \forall U\in \mathbb{U}, \rho \in \mathbb{S}\} \label{absolutelyseparableset}
\end{equation}
where $\mathbb{AS}$ denotes the set of absolutely separable states and $\mathbb{U}$ represents the set of all global unitary operations acting on the composite system. For example, the pure product states can always be transformed to an entangled state by the choice of a suitable global unitary, and hence they are not absolutely separable. The set of absolutely separable states is convex and compact \cite{PhysRevA.89.052304}. They were also characterized in a resource-theoretic paradigm where the absolutely separable states are free states, with the global unitaries and their convex mixtures acting as free operations \cite{PhysRevA.108.042402}.
 From the detection perspective, a necessary and sufficient criterion for absolute separability is given in \cite{PhysRevA.64.012316,PhysRevA.88.062330}. According to the criterion, a state $\rho \in \mathcal{D}(\mathbb{C}^2 \otimes \mathbb{C}^d)$ $\in \mathbb{AS}$ if and only if (iff) 
\begin{equation}
    \lambda_1^{\downarrow} -  \lambda_{2d-1}^{\downarrow} - 2 \sqrt{\lambda_{2d-2}^{\downarrow} \lambda_{2d}^{\downarrow}} \le 0 \label{abscriterion}
\end{equation}
where $\{\lambda_1^{\downarrow}, \lambda_2^{\downarrow}, ... , \lambda_{2d}^{\downarrow}\}$ are the eigenvalues of $\rho$ in decreasing order. This is also the condition for absoluteness of PPT states (states that remain PPT under the action of a global unitary), since absolutely PPT and absolute separability are equivalent for states $\in \mathcal{D}(\mathbb{C}^2 \otimes \mathbb{C}^d)$ \cite{PhysRevA.88.062330}, but in higher dimensions this is still an open question\cite{arunachalam2014absolute,song2024extreme,louvet2025nonequivalence}. Specifically, if $\mathbb{PPT}$ denotes the set of all PPT states, then the set of absolutely PPT states is defined as 
\begin{equation}
    \mathbb{APPT} = \{\rho : U \rho U^{\dagger} \in \mathbb{PPT} \hspace{2mm} \forall U\in \mathbb{U}, \rho \in \mathbb{PPT}\}, \label{absolutelyPPTset}
\end{equation} and $\mathbb{AS} = \mathbb{APPT}$ for $\rho \in \mathcal{D}(\mathbb{C}^2 \otimes \mathbb{C}^d)$. Unlike the absolutely separable states, the absolutely PPT states belonging to arbitrary dimensions are well characterized by means of necessary and sufficient conditions \cite{PhysRevA.76.052325}. However, such conditions are based on linear matrix inequalities, which are a function of the eigenvalues of the state. So, the detection of non-absolutely PPT states necessitates access to the full eigenvalue spectrum corresponding to the system by state tomography. But as the dimension of the system increases, the number of such linear inequalities increases exponentially \cite{johnston2014line}, and as the number of eigenvalues also increases, it becomes cumbersome to evaluate them by tomography \cite{paris2004quantum}.  

Moving beyond the regime of non-absolutely separating quantum states, another useful way of characterizing non-absolute separability is via the identification of specific type of quantum channels that are not absolutely separable. This is discussed explicitly in the following subsection.

\subsection{Absolutely separating channel}
Let $\boldsymbol{{\mathcal{E}}}$ denote the set of all bipartite quantum channels, i.e., $\mathcal{E}: \mathcal{M}_{d^2} \rightarrow \mathcal{M}_{d^2}$, where $\mathcal{E} \in \boldsymbol{{\mathcal{E}}}$. Let $\boldsymbol{\mathcal{E}}_{\mathbb{AS}} \subset \boldsymbol{\mathcal{E}}$ denote the set of absolutely separable channels. For a bipartite system, these are defined as follows \cite{filippov2017absolutely},
\begin{equation}
    \boldsymbol{\mathcal{E}}_{\mathbb{AS}} = \{\mathcal{E}_{\mathbb{AS}} : \mathcal{E}_{\mathbb{AS}}(\rho) \in \mathbb{AS},  \forall \, \rho \in \mathcal{D} (\mathbb{C}^d \otimes \mathbb{C}^d)\} \label{ABSseparatingmap}.
\end{equation}
Note that for an absolutely separating channel, Eq.\eqref{ABSseparatingmap} holds true for any arbitrary choice of input state. Channels which do not belong to the set as mentioned in Eq.\eqref{ABSseparatingmap} are non-absolutely separating. The authors in \cite{filippov2017absolutely} propose a necessary and sufficient criterion for absolutely separating quantum channels for a certain class of channels, called the \textit{covariant channels}.

Covariant quantum channels \cite{holevo1993note,holevo2005additivity} are characterized by their sensitivity to unitary rotations of input states. Precisely, the set of bipartite covariant quantum channels, $\boldsymbol{\mathcal{E}}_{cov} (\subset  \boldsymbol{\mathcal{E}}): \mathcal{M}_{d^2} \rightarrow \mathcal{M}_{d^2}$ are defined as
\begin{equation}
     \boldsymbol{\mathcal{E}}_{cov} = \{\mathcal{E}_{cov} : U \mathcal{E}_{cov}(\rho) U^{\dagger} = \mathcal{E}_{cov} (U \rho {\color{red}U}^{\dagger}), \forall U \in \mathbb{U}\}  \label{covariantchannels}
\end{equation}
for any arbitrary input state $\rho$. Another set of channels that shares a close relation with absolutely separating channels consists of \textit{entanglement-annihilating channels} \cite{moravvcikova2010entanglement}.

Consider a bipartite quantum state $\rho$, then entanglement-annihilating channels $\boldsymbol{\mathcal{E}}_{an} (\subset \boldsymbol{\mathcal{E}}) : \mathcal{M}_{d^2} \rightarrow \mathcal{M}_{d^2}$ are defined as, 
\begin{equation} 
    \boldsymbol{\mathcal{E}}_{an} = \{ \mathcal{E}_{an} : \mathcal{E}_{an}(\rho) \in \mathbb{S}, \; \forall \rho \in \mathcal{D}(\mathbb{C}^{d} \otimes \mathbb{C}^{d})\}, \label{annihilatingchannels} 
\end{equation}

Using these definitions, we now proceed to analyze the necessary and sufficient condition for a channel to be absolutely separating.

\begin{observation} \label{observation1}
   A covariant channel \( \mathcal{E}_{\text{cov}} \) acting on a bipartite system is absolutely separating if and only if it is entanglement-annihilating.
\cite{filippov2017absolutely}.  \label{observation1} 
\end{observation}
For a detailed proof, interested readers are referred to \cite{filippov2017absolutely}. So, this observation is a way to detect non-absolute separability for covariant quantum channels.

However, it is noteworthy to mention that the characterization of non-absolute separability for arbitrary states and channels is based on whether the unitary transformed state is entangled or not. But, entanglement detection is an NP-hard problem in general \cite{even1980cryptocomplexity}, and hence the detection of non-absolutely separable states and non-absolutely separating channels is challenging. In this paper, we present an efficient way to detect non-absolute separability for states and channels of arbitrary dimensions using the concept of moments. For the sake of completeness, in the next subsection, we review the use of moments in the context of bipartite entanglement detection.
\subsection{Moment criteria}
 For a bipartite quantum state $\rho_{AB}$, the $n$-th order partial transpose (PT) moments \cite{gray2018machine,elben2020mixed,yu2021optimal,Neven2021,mallick2024assessing,mallick2025efficient,mallick2025higher,mukherjee2025detecting,chakrabarty2025probing,aggarwal2024entanglement,kumar2025construction,aggarwal2024theoretical} are defined as
\begin{equation}
p_n := \mathrm{Tr}[(\rho_{AB}^{T_A})^n], \label{PTmoments}
\end{equation}
for $n \in \mathbb{N}$. These moments serve as useful tools for probing quantum correlations, particularly in the context of many-body systems and relativistic quantum field theory. The spectrum ${\lambda_i}$ of the partially transposed state $\rho^{T_A}_{AB}$ corresponds to the roots of its characteristic polynomial, given by
\begin{equation}
\text{Det}(\rho^{T_A}_{AB} - \lambda I) = \sum_{n} a_{n} \lambda^{n},
\end{equation}
where the coefficients $a_n$ can be expressed as functions of the PT moments $p_n$ introduced in Eq.~\eqref{PTmoments}. Thus, the PT moments contain essential spectral information of $\rho^{T_A}_{AB}$, enabling a characterization of its eigenvalues through quantities that can be accessed experimentally. 

In \cite{elben2020mixed}, Elben et al. introduced a criterion for entanglement detection based on the first three moments of the partial transpose. Their criterion asserts that for any bipartite quantum state $\rho_{AB}$ that is positive under partial transposition (PPT), the inequality ${p^2_2} \leq p_3 p_1$ must be satisfied. A violation of this inequality indicates that the state is non-positive under partial transposition (NPT), thereby certifying the presence of entanglement. This criterion is commonly known as the $p_3$-PPT criterion. 

Yu et al.~\cite{yu2021optimal}  generalized this criterion by incorporating higher-order moments, leading to a stronger entanglement detection criterion than the earlier $p_3$-PPT criterion. Specifically, they introduced the notion of Hankel matrices constructed from the sequence of partial transpose moments $\mathbf{p} = (p_1, p_2, \dots, p_n)$. These matrices, denoted by $H_n(\mathbf{p})$ are symmetric $(n+1) \times (n+1)$ matrices with entries defined as
\begin{equation}
[H_n(\mathbf{p})]_{ij} := p_{i+j+1}, \quad \text{for } i,j \in {0, 1, \dots, n}. \label{Hankelmatrices}
\end{equation}

The first few Hankel matrices take the explicit forms:
\begin{equation}
    H_1 = \begin{pmatrix}
p_1 & p_2   \vspace{0.2cm}\\ 
p_2 & p_3   
\end{pmatrix} 
\hspace{0.1cm} \text{and} \hspace{0.1cm}  H_2 = \begin{pmatrix}
p_1 & p_2 & p_3  \vspace{0.2cm}\\ 
p_2 & p_3 & p_4  \vspace{0.2cm}\\
p_3 & p_4 & p_5      \label{secondHankelmatrix}
\end{pmatrix}
\end{equation} 
Based on this structure of Hankel matrices, a generalized necessary condition for the separability  is given by:
    \begin{equation}
        \det[H_{n}(\mathbf{p})] \ge 0 . \label{Hankelmatrixcondition}
    \end{equation} 
   These PT moments can be efficiently estimated in a real experiment through shadow tomography, which prevents the need for full quantum state tomography, thereby substantially reducing resource requirements \cite{aaronson2018shadow,aaronson2019gentle,huang2020predicting,elben2020mixed}. These PT moments-based methods provide a significant advantage in NISQ devices and many-body quantum systems, where a single qubit can act as a control, allowing multiple partial transpose moments to be extracted from the same experimental data, unlike approaches that rely on random global unitaries for randomized measurements \cite{huang2020predicting}.

With this motivation, we now proceed to develop a moment-based framework for detecting non-absolutely separable states. We begin by defining a sequence of moments $(s_n)$, and depending on it, we formulate a criterion to detect non-absolutely separable states.

\section{Moment-based detection of non-absolutely separable states and non-absolutely PPT states} \label{s3}

\subsection{Detection of non-absolute separability}
\begin{definition} \label{definition1}
Let $\Lambda:\mathcal{M}_d\rightarrow\mathcal{M}_d$ be a linear, positive but not completely positive map, and let $\mathcal{U} \in \mathbb{U}$ denote a global unitary operator. We define the $n$-th order moments $s_n$ corresponding to the map $\Lambda$ as follows:
\begin{equation} \label{moments}
s_n := \text{Tr}[S_{\Lambda}^n],
\end{equation}
where
\begin{equation} \label{def_S}
S_{\Lambda} = \frac{( \mathrm{id}_A \otimes \Lambda)(\mathcal{U} \, \rho_{AB} \, \mathcal{U}^{\dagger})}{\Tr[( \mathrm{id}_A \otimes \Lambda)(\mathcal{U} \, \rho_{AB} \, \mathcal{U}^{\dagger})]},
\end{equation}
and $n$ is a positive integer.
\end{definition}
With the above definition, we now proceed to formulate our criterion for the detection of non-absolutely separable states.\\

\begin{theorem} \label{theorem1}
Let $\rho_{AB}$ be an absolutely separable state. Then, for all global unitary operators $\mathcal{U} \in \mathbb{U}$, the following inequality holds:
\begin{equation}
s_2^2 \leq s_3 \label{6}
\end{equation}
where $s_2$ and $s_3$ are defined as in equation~\eqref{moments}.
\end{theorem}

\proof  Let $\rho_{AB} \in \mathbb{AS}$, i.e., for any global unitary $\mathcal{U}$, the state $\mathcal{U} \, \rho_{AB} \, \mathcal{U}^\dagger$ remains separable. Given a positive but not completely positive map $\Lambda$, consider the operator $S_{\Lambda}$ given by Eq.\eqref{def_S}.

According to the properties of positive maps \cite{HORODECKI19961}, it follows that $S_{\Lambda}$ is a positive semidefinite operator with unit trace. Let us define Schatten-$p$ norms for $p \ge 1$ on the positive semidefinite operator $S_{\Lambda}$ as
 \begin{equation}
    ||S_{\Lambda}||_{p} := (\sum_{i=1}^{d}{|\chi_i|^p})^{\frac{1}{p}}=(\text{Tr}[|S_{\Lambda}|^p])^{\frac{1}{p}}   \label{schattenp}
\end{equation}
where $S_{\Lambda}$ admits the spectral decomposition $S_{\Lambda} = \sum_{i=1}^{d} \chi_i \ket{x_i}\bra{x_i}$. Moreover, the $\ell_p$ norm of the eigenvalue vector $\mathlarger{\chi}$ corresponding to the operator $S_{\Lambda}$ is given by
  \begin{equation}
     || \mathlarger{\chi}||_{l_p} := (\sum_{i=1}^{d}{|\chi_i|^p})^{\frac{1}{p}} \label{lpnorm}
 \end{equation}
 where ${\{\chi_i \}_{{i=1}}^{d}}$ denotes the eigenvalue spectrum of  $S_{\Lambda}$.

Let $p,q \ge 1$ satisfy $\frac{1}{p}+\frac{1}{q}=1$. For any vectors
$v,w \in \mathbb{C}^{d}$, Hölder's inequality states that
\begin{equation}
    \big|\langle v,w\rangle\big|
    = \Big|\sum_{i=1}^{d} \overline{v_i} w_i\Big|
    \le \|v\|_{p}\,\|w\|_{q}.
    \label{eq:holder}
\end{equation}

We consider the space of $ d\times d$ matrices as a finite-dimensional
inner-product space endowed with the inner product $ \langle A, B \rangle = \operatorname{Tr}(A^{\dagger}B)$.
 For a matrix
$S_{\Lambda}$, we define $\chi = \mathrm{vec}(S_{\Lambda}) \in \mathbb{C}^{d^2}$
as its column vectorization. Therefore,
\begin{equation}
    \langle \chi, \chi \rangle
    = \|\chi\|_{2}^{2}
    = \mathrm{Tr}(S_{\Lambda}^{\dagger} S_{\Lambda}) = \mathrm{Tr}[S_{\Lambda}^{2}], \label{19}
\end{equation}
since $S_{\Lambda}$ is Hermitian. Now from Eq.~\eqref{eq:holder} and Eq.\eqref{19} with $p = 3$, $q = \frac{3}{2}$, and $v = w = \chi$,
we obtain
\begin{equation}
    \mathrm{Tr}[S_{\Lambda}^{2}]
    = \|\chi\|_{2}^{2}
    \le \|\chi\|_{3}\,\|\chi\|_{3/2}
    \label{eq:holder_application}
\end{equation}

 Note that the Cauchy-Schwarz inequality is obtained by putting $p=2$ and $q=2$ in H\"{o}lder's inequality defined in Eq.\eqref{eq:holder}. \\
 Now, 
 \begin{align}
&{||S_{\Lambda}||_{2}}^2=\text{Tr}[{S_{\Lambda}}^2] \nonumber \\ 
& ~~~~~ \overset{a}{\leq} ||S_{\Lambda}||_{3} ||\mathlarger{\chi}||_{\ell_{\frac{3}{2}}} \nonumber \\ 
&   ~~~~~ =  ||S_{\Lambda}||_{3} (\sum_{i=1}^{d}{|\chi_i|^{\frac{3}{2}}})^{\frac{2}{3}} \nonumber \\ 
&     ~~~~~ =  ||S_{\Lambda}||_{3} (\sum_{i=1}^{d}{{|\chi_i|}{|\chi_i|}^{\frac{1}{2}}})^{\frac{2}{3}} \nonumber \\ 
&     ~~~~~  \overset{b}{\leq}  ||S_{\Lambda}||_{3}  ((\sum_{i=1}^{d}{|\chi_i|^2})^{\frac{1}{2}} (\sum_{i=1}^{d}{|\chi_i|})^{\frac{1}{2}})^{\frac{2}{3}}\nonumber \\ 
&     ~~~~~ =||S_{\Lambda}||_{3}  {||S_{\Lambda}||_{2}}^{\frac{2}{3}} {||S_{\Lambda}||_{1}}^{\frac{1}{3}}
 \label{applyingcauchy}
\end{align}
where (a) follows from Eq.~\eqref{eq:holder_application}, and (b) follows from the Cauchy–Schwarz inequality.
Taking $3$rd power of Eq.~\eqref{applyingcauchy}, we get
\begin{equation}
   {||S_{\Lambda}||_{2}}^4 \le  {||S_{\Lambda}||_{3}}^3 ||S_{\Lambda}||_{1} . \label{applying3rdpower}
\end{equation}
As $S_{\Lambda}$ is trace-normalized, i.e., $\text{Tr}(S_{\Lambda}) = 1$, its trace norm satisfies $||S_{\Lambda}||_1 = 1$. Putting this into Eq.~\eqref{applying3rdpower}, we obtain the following inequality:
\begin{equation}
   {||S_{\Lambda}||_{2}}^4 \le  {||S_{\Lambda}||_{3}}^3  \label{normproof}
\end{equation}
{\it i.e.},
\begin{equation}
   {s_2}^2 \leq s_3 \label{theorem1proof}
\end{equation}
which completes the proof. \qed \\

The above theorem implies that the condition stated in Eq.~\eqref{6} is a necessary criterion for a state $\rho_{AB}$ to be absolutely separable. Therefore, any violation of this condition serves as a sufficient indicator that $\rho_{AB} \notin \mathbb{AS}$. \\

We now utilize higher-order moments to introduce an extended criterion that effectively identifies states belonging to the set $\mathbb{S} - \mathbb{AS}$.\\
\\
\begin{theorem} \label{theorem2}
If a bipartite quantum state $\rho_{AB}$ is absolutely separable, then for all global unitary operators $\mathcal{U} \in \mathbb{U}$
\begin{equation}
      \det[H_{m}(s_{\Lambda})] \ge 0 \label{Hankel}.
    \end{equation}
\end{theorem}
\proof Let, $\rho_{AB} \in  \mathbb{AS}$ i.e. $\mathcal{U} \,\rho_{AB}\, {\mathcal{U}}^{\dagger} $ is separable for all global unitary $\mathcal{U} \in \mathbb{U}$. Now, consider a positive but not completely positive map $\Lambda$.

By the properties of positive maps \cite{HORODECKI19961}, it follows that $S_{\Lambda}$ given by Eq.\eqref{def_S} is a positive semidefinite operator with unit trace. Consequently, if ${\{\chi_i \}_{{i=1}}^{d}}$ denote its eigenvalues, we have $\chi_i \ge 0$ for all $i = 1, 2, \dots, d$.

Now, if $s_{\Lambda} = (s_1,s_2,....s_n)$ be the moment vector defined in Eq. \eqref{moments}, then the $(m+1) \times (m+1)$ Hankel matrices are given by the elements $[H_{m}(s_{\Lambda})]_{ij} = s_{i+j+1},$ with $i,j \in \{ 0,1, ..., m\}$. These Hankel matrices $H_{m}(s_{\Lambda})$ admit the following decomposition: 
  \begin{equation}
      H_{m}(s_{\Lambda}) = V_m D V_m^T
  \end{equation}
where,\begin{equation}
    V_m = \begin{pmatrix}
1 & 1 & ...&1  \vspace{0.2cm}\\ 
\chi_1 & \chi_2 & ...& \chi_d  \vspace{0.2cm}\\
...&...&...&...&\\
...&...&...&...&\\
\chi_1^m & \chi_2^m & ...& \chi_d^m      \label{secondHankelmatrix}
\end{pmatrix}
\end{equation} 
and \begin{equation}
    D = \begin{pmatrix}
 \chi_1& 0 & ...&0  \vspace{0.2cm}\\ 
0 & \chi_2 & ...& 0  \vspace{0.2cm}\\
...&...&...&...&\\
...&...&...&...&\\
0 & 0 & ...&   \chi_d  \label{secondHankelmatrix}
\end{pmatrix}.
\end{equation} 

Now, for an arbitrary vector $x=(x_1,...x_m,x_{m+1}) \in {\mathbb{R}}^{m+1}$, we have 
\begin{equation}
    x  H_{m}(s_{\Lambda}) x^T = x  V_m D V_m^T  x^T = y D y^T = \sum_{i=1}^d \chi_i {y_i}^2 \ge 0,
\end{equation}
where, $y= x  V_m = (y_1,y_2,...y_d) $ with $y_i= \sum_{j=1}^{m+1} x_j \hspace{0.1cm}{\chi_i}^{j-1}, \hspace{0.1cm} \text{ for } i=1,2,....d$. \\
Hence, $ x  H_{m}(s_{\Lambda}) x^T \ge 0$ which implies $ H_{m}(s_{\Lambda}) \ge 0$, {\it i.e.} $\det[H_{m}(s_{\Lambda})] \ge 0$. This completes the proof. \qed\\

Analogous to the Theorem \ref{theorem1}, the above result establishes that the condition in Eq.~\eqref{Hankel} is a necessary criterion for absolute separability. Consequently, any violation of this condition is sufficient to certify that the quantum state belongs to the set $\mathbb{S} - \mathbb{AS}$, i.e., it is not absolutely separable.

\subsubsection{Examples:} We now present three examples illustrating the utility of the above Theorems \ref{theorem1} and \ref{theorem2}.

\begin{example} \label{example1}
Let us consider a separable state in $\mathcal{D}(\mathbb{C}^2 \otimes \mathbb{C}^2)$ and a unitary operator defined as follows:
\begin{equation}
\rho_1 = \frac{1}{4} \begin{pmatrix}
1 & 0 & 0 & 1 \\
0 & 1 & 1 & 0 \\
0 & 1 & 1 & 0 \\
1 & 0 & 0 & 1
\end{pmatrix}
\end{equation}
\begin{equation}
\mathcal{U}_1 = \frac{1}{\sqrt{2}} \begin{pmatrix}
1 & 0 & 0 & 1 \\
0 & \sqrt{2} & 0 & 0 \\
0 & 0 & \sqrt{2} & 0 \\
-1 & 0 & 0 & 1
\end{pmatrix}
\end{equation}
\end{example}
Applying our criterion from Theorem \ref{theorem1} with the choice $\Lambda = \mathcal{T}$, where $\mathcal{T}$ denotes the standard transposition map, we find that the condition $s_2^2 - s_3 = 0$ holds. Therefore, Theorem \ref{theorem1} fails to detect the state $\rho_1$. \\
However, upon applying our proposed criterion from Theorem \ref{theorem2} and using $\Lambda = \mathcal{T}$, we find that $\det[H_2(s_\mathcal{T})]$ is not positive semidefinite. This violation confirms that Theorem \ref{theorem2} successfully detects $\rho_1$.

Moreover, this state can also be detected by employing the reduction map $(\mathcal{R})$ in place of the transposition map $(\mathcal{T})$ in Eq.~\eqref{def_S}. Recall that the action of the reduction map~\cite{horodecki1999reduction} $\mathcal{R}:\mathcal{M}_d \rightarrow \mathcal{M}_d$ is defined as
\begin{equation}
    \mathcal{R}(\rho) = \Tr(\rho)\, I - \rho.
\end{equation}
Analogous to the transposition map, we observe that Theorem~\ref{theorem1} fails to detect this state when $\Lambda = \mathcal{R}$. In contrast, Theorem~\ref{theorem2}, which incorporates moments up to the fifth order, successfully detects it.
\begin{example} \label{example2} 
Consider a separable state in $\mathcal{D}(\mathbb{C}^2 \otimes \mathbb{C}^4)$ and a unitary operator defined as follows:
\begin{equation}
\rho_2 =  \begin{pmatrix}
\frac{1}{4} & 0 & \frac{1}{4} & 0 &0&0&0&0 \\
0 & 0 & 0 & 0 & 0& 0& 0& 0 \\
\frac{1}{4} & 0 & \frac{1}{4} & 0 &0&0&0&0 \\
0 & 0 & 0 & 0 & 0& 0& 0& 0\\
0 & 0 & 0 & 0 & 0& 0& 0& 0\\
0 & 0 & 0 & 0 & 0& \frac{1}{4}& 0& \frac{1}{4}\\
0 & 0 & 0 & 0 & 0& 0& 0& 0\\
0 & 0 & 0 & 0 & 0& \frac{1}{4}& 0& \frac{1}{4}
\end{pmatrix}
\end{equation}
\begin{equation}
\mathcal{U}_2 = \frac{1}{\sqrt{2}} \begin{pmatrix}
1 & 0 & \frac{1}{4} & 0 &0&0&0&1 \\
0 & {\sqrt{2}} & 0 & 0 & 0& 0& 0& 0 \\
0 & 0 & {\sqrt{2}} & 0 &0&0&0&0 \\
0 & 0 & 0 & {\sqrt{2}} & 0& 0& 0& 0\\
0 & 0 & 0 & 0 & {\sqrt{2}}& 0& 0& 0\\
0 & 0 & 0 & 0 & 0& {\sqrt{2}}& 0& 0\\
0 & 0 & 0 & 0 & 0& 0& {\sqrt{2}}& 0\\
-1 & 0 & 0 & 0 & 0& 1& 0& 1
\end{pmatrix}
\end{equation}
\end{example}
Utilizing the criterion presented in Theorem \ref{theorem1} with the choice $\Lambda = \mathcal{T}$, where $\mathcal{T}$ denotes the transposition map, we find that the condition $s_2^2-s_3 > 0$ is satisfied. This confirms that Theorem \ref{theorem1} successfully identifies the non-absolute separability of the state $\rho_2$.
Note that this state is also detected by using moments of the reduction map, i.e., by putting $\Lambda = \mathcal{R}$ in Eq.\eqref{def_S}, and using Theorem~\ref{theorem1}.
\begin{example} \label{example3}
Consider the isotropic state $\rho_3 \in \mathcal{D}(\mathbb{C}^3 \otimes \mathbb{C}^3)$, given by
\begin{equation}
    \rho_3= p \ket{\phi^+}\bra{\phi^+} + \frac{(1-p)}{9} I_9
\end{equation}
where $\ket{\phi^+} = \frac{1}{\sqrt{3}} (\ket{00} +\ket{11} +\ket{22})$ is the maximally entangled state, $p \in [0,1]$ is the convex coefficient and $I_9$ is the $9$-dimensional Identity matrix.
\end{example}
This state is separable for $p \in [0, \frac{1}{4}]$ \cite{PhysRevA.72.052331}. Consider a transformation of this state via a global unitary given by 
\begin{equation}
\begin{split}
    \mathcal{U}_3 = & I_9 - \frac{\sqrt{2}-1}{\sqrt{2}} (\ket{00}\bra{00} + \ket{22}\bra{22}) \\ & + \frac{1}{\sqrt{2}} (\ket{00}\bra{22} - \ket{22}\bra{00}).
    \end{split}
\end{equation}
We find that the moments up to third order (the criterion of Theorem \ref{theorem1}) are not able to detect the state $\mathcal{U}_3 \, \rho_3  \,\mathcal{U}_3^{\dagger}$. However, the criterion of Theorem \ref{theorem2} involving higher order moments are able to detect this state for $p \ge 0.203$. Specifically, for $p \in [0.203, 0.25]$, $\det[H_2(s_\mathcal{T})]$ is not positive semi-definite. Hence $\rho_3 \notin \mathbb{AS}$ for $p \in [0.203, 0.25]$.

\subsection{Detection of non-absolutely PPT states}
As mentioned in the introduction, a closely related problem is the detection of non-absolutely PPT states. In this subsection, we discuss the detection of non-absolutely PPT states using moments. However, before proceeding further, we provide a brief overview of the definition and detection of PPT states for the sake of completeness.

    A bipartite quantum state $\rho_{AB}$ is PPT if and only if $(\mathrm{id}_A \otimes T)(\rho_{AB}) \geq 0$, where $T$ denotes the transposition map. This definition can, in fact, be extended to encompass a broader class of maps known as \textit{decomposable maps} \cite{stormer1963positive}, of which the transposition map is a particular example. Note that, If a map $\Lambda$ can be written as $\Lambda = \Lambda_1 + \Lambda_2 \circ T$
where $\Lambda_1, \Lambda_2$ are completely positive maps, then $\Lambda$ is said to be decomposable.

Owing to the structural properties of PPT states, decomposable maps are incapable of detecting such states. Consequently, the definition of PPT remains valid even when the transposition map $T$ is replaced by an arbitrary decomposable map $\Lambda$. 

In contrast, non-absolutely PPT states are characterized by their ability to transform into NPT states by the action of suitable global unitaries. Decomposable maps can detect NPT states. Since detecting a non-absolutely PPT state is equivalent to detecting the corresponding global unitary transformed NPT state, using decomposable maps suffices for our purpose. Hence, unless stated otherwise, we shall restrict ourselves to decomposable maps throughout this subsection.

\begin{definition} \label{Definition2}
    Let $\tilde{\Lambda}:\mathcal{M}_d\rightarrow\mathcal{M}_d$ be a linear, positive but not completely positive, decomposable map, and let $\mathcal{U} \in \mathbb{U}$ denote a global unitary operator. We define the $n$-th order moments $r_n$ corresponding to the map $\tilde{\Lambda}$ as follows:
\begin{equation} \label{moments1}
r_n := \text{Tr}[R_{\tilde{\Lambda}}^n]
\end{equation}
where
\begin{equation} \label{def_R}
R_{\tilde{\Lambda}}= \frac{( \mathrm{id}_A \otimes \tilde{\Lambda})(\mathcal{U} \, \rho_{AB} \, \mathcal{U}^{\dagger})}{\text{Tr}[( \mathrm{id}_A \otimes \tilde{\Lambda})(\mathcal{U} \, \rho_{AB} \, \mathcal{U}^{\dagger})]},
\end{equation}
and $n$ is a positive integer.
\end{definition}
With the above definition, we now proceed to formulate our criterion for the detection of non-absolutely PPT states.\\

\begin{theorem} \label{theorem3}
Let $\rho_{AB}$ be an absolutely PPT state. Then, for all global unitary operators $\mathcal{U} \in \mathbb{U}$, the following inequality holds:
\begin{equation}
r_2^2 \leq r_3 \label{7}
\end{equation}
where $r_2$ and $r_3$ are defined in equation~\eqref{moments1}.
\end{theorem}

\proof  Let $\rho_{AB} \in \mathbb{APPT}$, i.e., for any global unitary $\mathcal{U}$, the state $\mathcal{U} \, \rho_{AB} \, \mathcal{U}^\dagger$ remains PPT. Given a positive but not completely positive decomposable map $\tilde{\Lambda}$, consider the operator $R_{\tilde{\Lambda}}$ given by Eq.\eqref{def_R}.

According to the properties of positive maps \cite{HORODECKI19961}, it follows that $R_{\tilde{\Lambda}}$ is a positive semidefinite operator with unit trace. Rest of the proof then follows similarly to Theorem \ref{theorem1}, replacing $S_{\Lambda}$ by $R_{\tilde{\Lambda}}$. \qed\\

We now utilize higher-order moments to introduce an extended criterion that effectively identifies states belonging to the set $\mathbb{PPT} - \mathbb{APPT}$.\\
\\
\begin{theorem} \label{theorem4}
If a bipartite quantum state $\rho_{AB}$ is absolutely PPT, then for all global unitary operators $\mathcal{U} \in \mathbb{U}$
\begin{equation}
       \det[H_{m}(\mathbf{R}_{\tilde{\Lambda}})] \ge 0 . \label{Hankelmatrix}
    \end{equation}
Here, $[H_{m}(\mathbf{R}_{\Lambda})]_{ij} = r_{i+j+1}$ for $i,j \in \{0,1,...,m\}$, $m \in \mathbb{N}$ and $r_i$, $i= 1,2,..,n$  are the i-th moments defined in Eq.~\eqref{moments1}.
\end{theorem}
\proof Let, $\rho_{AB} \in  \mathbb{APPT}$ i.e. $\mathcal{U} \,\rho_{AB}\, {\mathcal{U}}^{\dagger} $ is PPT for all global unitary $\mathcal{U} \in \mathbb{U}$. Now, consider a positive but not completely positive map $\tilde{\Lambda}$. 

By the properties of positive maps \cite{HORODECKI19961}, it follows that $R_{\tilde{\Lambda}}$ given by Eq.\eqref{def_R} is a positive semidefinite operator with unit trace. Consequently, if ${\{\chi_i \}_{{i=1}}^{d}}$ denote its eigenvalues, we have $\chi_i \ge 0$ for all $i = 1, 2, \dots, d$. 

Now, if $\mathbf{R}_{\tilde{\Lambda}} = (r_1,r_2,....r_n)$ be the moment vector defined in Eq. \eqref{moments1}, then the $(m+1) \times (m+1)$ Hankel matrices are given by the elements $[H_{m}(\mathbf{S}_{\tilde{\Lambda}})]_{ij} = s_{i+j+1},$ with $i,j \in \{ 0,1, ..., m\}$. The remainder of the proof then follows analogously to Theorem \ref{theorem2}, replacing $\mathbf{S}_{\Lambda}$ by $\mathbf{R}_{\tilde{\Lambda}}$. \qed
\begin{theorem} \label{theorem9}
(\textit{Sufficient condition}) : If for all global unitary operators $\mathcal{U} \in \mathbb{U}$ and for all linear, positive but not completely positive decomposable map $\tilde{\Lambda}$, a bipartite quantum state $\rho_{AB} \in \mathcal{D}(\mathbb{C}^d \otimes \mathbb{C}^d)$ satisfies
\begin{equation}
      r_2 \le \frac{1}{d^2-1}
    \end{equation}
where, $r_2$ is defined in Eq. \eqref{moments1}, then the state $\rho_{AB}$ is absolutely PPT \cite{mehta1989matrix}.
\end{theorem}
\proof  Let a bipartite quantum state $\rho_{AB} \in \mathcal{D}(\mathbb{C}^d \otimes \mathbb{C}^d)$ satisfy
\begin{equation}\label{condition}
      r_2 \le \frac{1}{d^2-1}
    \end{equation}
for all global unitary operators $\mathcal{U} \in \mathbb{U}$.  This implies that
\begin{equation}
    \mathrm{Tr}[R_{\tilde{\Lambda}}^2] \le \frac{1}{d^2-1},
\end{equation}
where $R_{\tilde{\Lambda}}$ is given by Eq.\eqref{def_R}. 

To prove that $\rho_{AB}$ is absolutely PPT, we must show that
\[
(\mathrm{id}_A \otimes \tilde{\Lambda})(\mathcal{U} \rho_{AB} \mathcal{U}^\dagger) \ge 0
\]
for all $\mathcal{U} \in \mathbb{U}$ and arbitrary map $\tilde{\Lambda}$. We proceed by contradiction.

Suppose there exists a unitary $\mathcal{U} \in \mathbb{U}$ such that
\begin{equation}\label{initialassumption}
    (\mathrm{id}_A \otimes \tilde{\Lambda})(\mathcal{U} \rho_{AB} \mathcal{U}^\dagger) \ngeq 0,
\end{equation}
i.e., the operator has at least one negative eigenvalue. Let $\{\chi_i\}_{i=1}^{d^2}$ denote the eigenvalues of this operator, and assume without loss of generality that $\chi_1 < 0$.

Now,
\begin{align*}
    \mathrm{Tr}(R_{\tilde{\Lambda}}) &= \sum_{i=1}^{d^2} \chi_i \\
    & = \chi_1 + \sum_{i=2}^{d^2} \chi_i \\
    & <  \sum_{i=2}^{d^2} \chi_i  \hspace{0.4cm}[\text{Since,} \hspace{0.1cm}\chi_1 < 0]\\
    & \overset{a} \le \left(\sum_{i=2}^{d^2} \chi_i^2\right)^{1/2}
\left(\sum_{i=2}^{d^2} 1^2\right)^{1/2} \\
    &  = \sqrt{d^2-1} \left( \sum_{i=2}^{d^2} \chi_i^2 \right)^{1/2} \\ 
    & < \sqrt{d^2-1} \left( \sum_{i=1}^{d^2} \chi_i^2 \right)^{1/2} = \sqrt{d^2-1} \, [\large\mathrm{Tr}(R_{\tilde{\Lambda}}^2)\large]^{\frac{1}{2}}
\end{align*}
where ($a$) follows from the Cauchy–Schwarz inequality \cite{bhatia2013matrix}. Therefore,
 \begin{equation}
     \mathrm{Tr}(R_{\tilde{\Lambda}})  < \sqrt{d^2-1} \, [\mathrm{Tr}(R_{\tilde{\Lambda}}^2)]^{\frac{1}{2}}
 \end{equation}
Since, $\mathrm{Tr}(R_{\tilde{\Lambda}}) =1$, hence we obtain 
 \begin{equation}
     1  < \sqrt{d^2-1} \, [\mathrm{Tr}(R_{\tilde{\Lambda}}^2)]^{\frac{1}{2}}
 \end{equation}
Squaring both sides, we obtain 
\begin{equation}
    \mathrm{Tr}(R_{\tilde{\Lambda}}^2) > \frac{1}{d^2-1}
\end{equation}
i.e., \begin{equation}
    r_2 > \frac{1}{d^2-1}
\end{equation}

Thus, our initial assumption [Eq.\eqref{initialassumption}] must be false, and it follows that for all $\mathcal{U} \in \mathbb{U}$,
\[
(\mathrm{id}_A \otimes \tilde{\Lambda})(\mathcal{U}\, \rho_{AB}\, \mathcal{U}^\dagger) \ge 0,
\]
for all linear, positive but not completely decomposable positive maps, i.e., $\rho_{AB}$ is absolutely PPT. \qed\\

\textbf{Note:} The criterion established in Theorem~\eqref{theorem9} serves as a sufficient condition for a quantum state to be absolutely separable in the restricted dimension \(\mathbb{C}^2 \otimes \mathbb{C}^d\). Therefore, any state that is not absolutely separable in this dimension necessarily fails to satisfy the condition stated in Theorem~\eqref{theorem9}.

\begin{example} \label{example4}
Consider the state
\begin{equation}
    \rho_4 = p\, \rho_b + \frac{1-p}{9} \, I_9,
\end{equation}
where \begin{equation}
    \rho_b = \frac{2}{7} \ket{\phi^+} \bra{\phi^+}
           + \frac{b}{7} \, \sigma_{+}
           + \frac{5-b}{7} \, \sigma_{-},
\end{equation}
with $\ket{\phi^+} = \frac{1}{\sqrt{3}} (\ket{00} + \ket{11} + \ket{22})$, $\sigma_{+} = \frac{1}{3} \left( \ket{01}\bra{01} + \ket{12}\bra{12} + \ket{20}\bra{20} \right)$, and $\sigma_{-} = \frac{1}{3} \left( \ket{10}\bra{10} + \ket{21}\bra{21} + \ket{02}\bra{02} \right)$.
\end{example}

\begin{figure}[ht]
\includegraphics[width=.45\textwidth]{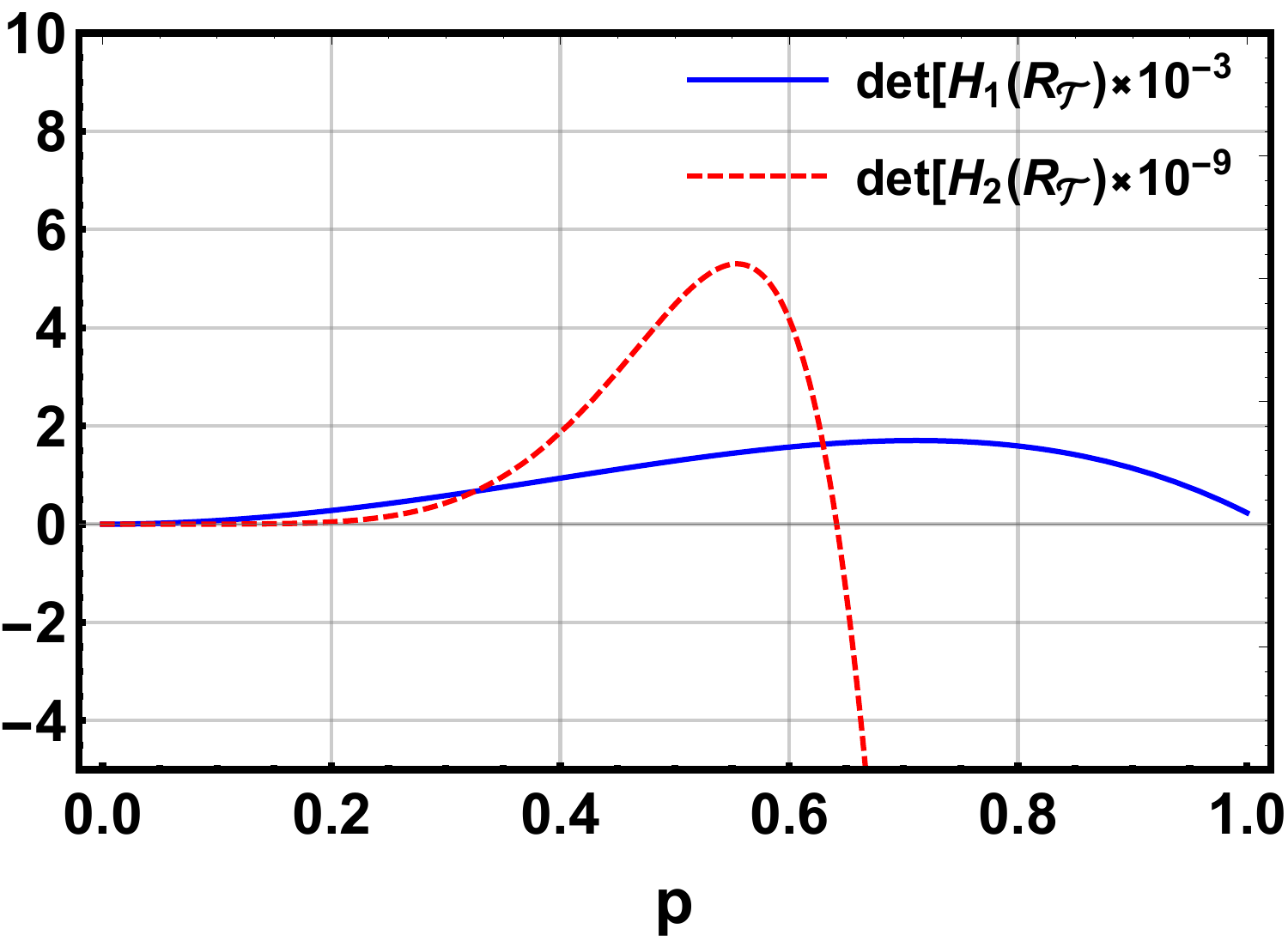} 
\caption{Detection of the non-absolutely PPT state \( \rho_4 \) using Theorem~\ref{theorem3} and Theorem~\ref{theorem4}. The plot shows that \( \det[H_{2}(\mathbf{R}_{\mathcal{T}})] < 0 \) for \( p > 0.65 \).}\label{fig2}

\centering
\end{figure}

It is known that $\rho_b$ is PPT for $1 \leq b \leq 4$.  
Now, consider applying to this state a global unitary transformation of the form
\begin{equation}\label{eq:Uphi}
\begin{aligned}
U_4(\phi_{1}, \phi_{2}) 
&= \cos\phi_{1} \, \big[ \sigma_{x} \otimes \sigma_{y} \otimes \sigma_{z} \big] \\
&\quad + \sin\phi_{1} \sin\phi_{2} \, \big[ \sigma_{y} \otimes \sigma_{z} \otimes \sigma_{x} \big] \\
&\quad + \sin\phi_{1} \cos\phi_{2} \, \big[ \sigma_{z} \otimes \sigma_{x} \otimes \sigma_{y} \big]  \oplus \begin{bmatrix} 1 \end{bmatrix} 
\end{aligned}
\end{equation}
where $\begin{bmatrix} 1 \end{bmatrix}$ denotes a $1 \times 1$ matrix with the single entry $1$. From \cite{patra2021efficient}, it follows that for \( b = 1.5 \) and parameters \(\phi_{1} = \frac{\pi}{18}\) and \(\phi_{2} = \frac{5\pi}{6}\), the global unitary \( U_4\!\left(\frac{\pi}{18}, \frac{5\pi}{6}\right) \) transforms the state \( \rho_4\) into a non-positive partial transpose (NPT) state whenever \( p > 0.6 \). This demonstrates that \( \rho_4\) is not absolutely PPT for \( p > 0.6 \). Our moment-based condition in Theorem~\ref{theorem3} fails to detect the non-absolutely PPT nature of \( \rho_4\) in this range. However, Theorem~\ref{theorem4} successfully detects \( \rho_4\) i.e., \( \det[H_{2}(\mathbf{R}_{\mathcal{T}})] < 0 \) for \( p > 0.65 \), , which is evident from Fig.~\ref{fig2}.

\section{Detection of non-absolutely separating covariant quantum channels} \label{s4}
In this section, we aim to detect non-absolutely separating covariant channels using our moment-based criteria.\\

\begin{definition} \label{definition3}
    Let $\Lambda:\mathcal{M}_d \rightarrow \mathcal{M}_d$ be a linear, positive, but not completely positive map. We define the $n$-th order moment for detecting non-absolutely separating channels as
\begin{equation} \label{moments3}
q_n := \text{Tr}[Q_{\Lambda}^n],
\end{equation}
where
\begin{equation} \label{def_Q}
Q_{\Lambda} = \frac{(\mathrm{id}_{B_1} \otimes \Lambda)(\mathcal{E}_{an}(\omega_{B_1B_2}))}{\text{Tr}[(\mathrm{id}_{B_1} \otimes \Lambda)(\mathcal{E}_{an}(\omega_{B_1B_2}))]},
\end{equation}
and $\omega_{B_1B_2} \in \mathcal{D}(\mathcal{H}_B)$ is an arbitrary bipartite input state. Here, $\mathcal{E}_{an}$ denotes an entanglement-annihilating channel as defined in Eq.~(\ref{annihilatingchannels}), and $n \in \mathbb{N}$ is a positive integer.
\end{definition} 
With this definition in place, we now proceed to establish our criterion for detecting non-absolutely separating channels.

\begin{theorem} \label{theorem8}
Let $\mathcal{E}_{an}$ be an entanglement annihilating channel acting on a bipartite system $B$. Then, for any positive, but not completely positive map $\Lambda$, and arbitrary input state $\omega_{B_1B_2}$, the following inequality holds:
\begin{equation} 
q_2^2 \leq q_3 \label{10}
\end{equation}
where $q_2$ and $q_3$ are defined as in equation~\eqref{moments3}.
\end{theorem}
\proof Let $\mathcal{E}_{an}$ be an entanglement annihilating channel acting on a bipartite system $B$. Then, for any positive, but not completely positive map $\Lambda$, consider the operator given by Eq.\eqref{def_Q}, for an arbitrary input state $\omega_{B_1B_2}$. From the definition of an entanglement-annihilating channel, we obtain that $Q_{\Lambda}$ is a positive semidefinite operator with unit trace. The remainder of the proof follows similarly from Theorem \eqref{theorem1} by replacing $S_{\Lambda}$ by  $Q_{\Lambda}$.  \qed\\

Since from Ref.\cite{filippov2017absolutely}, we know that a covariant channel is absolutely separating if and only if it is entanglement annihilating.  Therefore, our criterion in Theorem \eqref{theorem8} provides a necessary condition for a covariant channel to be absolutely separating. Therefore, any violation of Eq.~\eqref{10} implies that the channel is not absolutely separating.  
\subsection{Example}
\begin{example}
   Consider the two-qubit depolarizing channel $ ( {\mathcal{E}_{dep}^{(2)}})$, whose action is given by \cite{nielsen2010quantum},
\begin{equation} \label{depolarzing}
    {\mathcal{E}_{dep}^{(2)}}(\rho) = p \rho + \frac{1-p}{4} \text{Tr}(\rho) I_4
\end{equation}
where, $p\in [0,1]$. \end{example} 

The global qubit depolarizing channel on an arbitrary two-qubit input state acts as follows:
\begin{equation}
     \omega_{out} = {\mathcal{E}_{dep}^{(2)}}  (\omega_{B_1 B_2}) = p \omega_{B_1 B_2} + \frac{1-p}{4}I_4 \\
\end{equation}

Now, to conclude that ${\mathcal{E}_{dep}^{(2)}}$ is an entanglement-annihilating channel, we need to verify for all input states $\omega_{B_1B_2}$. Since the set of separable states forms a convex set, it suffices to check only for pure states. Any two-qubit pure state can be written in the Schmidt decomposition form as :
\begin{equation}
    \ket{\psi} = \sum_{j=0}^1 \sqrt{q_j} \ket{\phi_j} \otimes \ket{\widetilde{\phi}_j}
\end{equation}
where $\ket{\phi_j}$ and $\ket{\widetilde{\phi}_j}$ are orthonormal bases for the two subsystems on Bob's side. Therefore, the output state takes the form 

  \begin{equation} \label{FBoutmat}
\omega_{\text{out}} = 
\begin{bmatrix}
pq_0+\frac{1-p}{4} & 0 & 0 & p\sqrt{q_0 q_1} \\
0 & \frac{(1 - p)}{4} & 0 & 0 \\
0 & 0 & \frac{(1 - p)}{4} & 0 \\
p\sqrt{q_0 q_1} & 0 & 0 & pq_1+\frac{1-p}{4}
\end{bmatrix}
\end{equation}
where $q_{0}\in[0,1]$ and $q_{0}=1-q_{1}$. By applying the criterion established in Theorem~\ref{theorem8} with the choice $\Lambda = \mathcal{T}$, where $\mathcal{T}$ denotes the transposition map, we observe that the condition $q_2^2 - q_3 > 0$ holds for $p > \frac{1}{3}$. This result demonstrates that Theorem~\ref{theorem8} effectively identifies the parameter regime in which the channel ${\mathcal{E}_{dep}^{(2)}}$ is not entanglement annihilating. Hence, from the equivalence relation presented as discussed in Observation \ref{observation1}\cite{filippov2017absolutely}, we can conclude that our proposed criterion is capable of detecting the region where the channel ${\mathcal{E}_{dep}^{(2)}}$ is not absolutely separating.
\begin{example} 
   Consider the global qutrit depolarizing channel $ ( {\mathcal{E}_{dep}^{(3)}})$, whose action is given by ,
\begin{equation} \label{depolarzing}
    {\mathcal{E}_{dep}^{(3)}}(\rho) = p {\rho} + \frac{1-p}{9} \text{Tr}(\rho) I_9
\end{equation}
where, $p\in [0,1]$.\end{example} 
The global qutrit depolarizing channel on an arbitrary two-qutrit input state acts as follows:
\begin{equation}
     \eta_{out} = {\mathcal{E}_{dep}^{(3)}}  ({\eta_{B_1 B_2}}) = p {\eta_{B_1 B_2}} + \frac{1-p}{9}I_9 
\end{equation}
 
Following a similar line of reasoning as in the previous example, to establish that $\mathcal{E}_{dep}^{(3)}$ is an entanglement-annihilating channel, it is sufficient to consider only pure input states. Without loss of generality, any two-qutrit pure state can be expressed in its Schmidt decomposition form as follows.

\begin{equation}
    \ket{\psi} = \sum_{j=0}^2 \sqrt{q_j} \ket{\phi_j} \otimes \ket{\widetilde{\phi}_j}
\end{equation}
where $\ket{\phi_j}$ and $\ket{\widetilde{\phi}_j}$ form orthonormal bases for the respective subsystems on Bob's side. Consequently, the resulting output state, denoted by $\eta_{out}$ is given by

\begin{equation} \label{FBoutmat}
\scalebox{0.80}{$
\left[
\begin{array}{ccccccccc}
pq_0+\frac{1-p}{9} & 0 & 0 & 0 & p\sqrt{q_0 q_1} & 0 & 0 & 0 & p\sqrt{q_0 q_2} \\[6pt]
0 & \frac{1-p}{9} & 0 & 0 & 0 & 0 & 0 & 0 & 0 \\[6pt]
0 & 0 & \frac{1-p}{9} & 0 & 0 & 0 & 0 & 0 & 0 \\[6pt]
0 & 0 & 0 & \frac{1-p}{9} & 0 & 0 & 0 & 0 & 0 \\[6pt]
p\sqrt{q_0 q_1} & 0 & 0 & 0 & pq_1+\frac{1-p}{9} & 0 & 0 & 0 & p\sqrt{q_1 q_2} \\[6pt]
0 & 0 & 0 & 0 & 0 & \frac{1-p}{9} & 0 & 0 & 0 \\[6pt]
0 & 0 & 0 & 0 & 0 & 0 & \frac{1-p}{9} & 0 & 0 \\[6pt]
0 & 0 & 0 & 0 & 0 & 0 & 0 & \frac{1-p}{9} & 0 \\[6pt]
p\sqrt{q_0 q_2} & 0 & 0 & 0 & p\sqrt{q_1 q_2} & 0 & 0 & 0 & pq_2+\frac{1-p}{9}
\end{array}
\right]
$}
\end{equation}

where $q_{0}, q_{1} \in [0,1]$ and satisfy the normalization condition $q_{0} + q_{1} = 1 - q_{2}$. Applying the criterion outlined in Theorem~\ref{theorem8} with the specific choice $\Lambda = \mathcal{T}$, where $\mathcal{T}$ represents the transposition map, we find that the inequality $q_2^2 - q_3 > 0$ is satisfied whenever $p > \frac{1}{4}$. This observation confirms that Theorem~\ref{theorem8} successfully captures the range of parameters for which the qutrit depolarizing channel $\mathcal{E}_{dep}^{(3)}$ fails to be entanglement-annihilating. Consequently, from the equivalence relation presented in Observation \ref{observation1}\cite{filippov2017absolutely}, we infer that the proposed criterion is also effective in identifying the region where the channel $\mathcal{E}_{dep}^{(3)}$ is not absolutely separating.

\section{Operational advantage of non-absolutely separable states} \label{s5}
In this subsection, we demonstrate an operational advantage of non-absolutely separable states by exploring their role in channel discrimination tasks.

The problem of quantum channel discrimination is fundamentally connected to the widely explored problem of quantum state discrimination \cite{helstrom1969quantum,piani2009all}. The evolution of quantum states is most generally modelled by quantum channels, which are completely positive and trace-preserving maps. In quantum information processing, the effectiveness of a channel often depends on the specific operational context—one channel may provide a distinct advantage over another for a particular task \cite{christandl2009postselection,macchiavello2013quantum,bae2019more}. This naturally leads to the significant problem of quantum channel discrimination.  Consider the simplest scenario of discriminating two quantum channels $\mathcal{E}_i$: $\mathcal{M}_d \rightarrow  \mathcal{M}_d$, for $i \in \{1,2\}$, each occurring with an equal a priori probability, and our goal is to guess the channel with minimal error probability. Given an input state $\rho$, the channels produce output states $\mathcal{E}_1(\rho)$ and $\mathcal{E}_2(\rho)$. Thus, the problem effectively reduces to distinguishing between these two resulting quantum states.

Probe-ancilla quantum state can enhance the success probability in distinguishing quantum channels \cite{kitaev1997quantum,d2001using}. Consider a bipartite state $\rho_{AB} \in \mathcal{D}(\mathbb{C}^d \otimes \mathbb{C}^d)$ prepared as the input, where subsystem \( B \) is subjected to one of two quantum channels, \( \mathcal{E}_1 \) or \( \mathcal{E}_2 \). The corresponding output states are
\begin{equation}
    \rho_{AB}^1 = (\mathrm{id}_A \otimes \mathcal{E}_1)(\rho_{AB}), \quad 
    \rho_{AB}^2 = (\mathrm{id}_A \otimes \mathcal{E}_2)(\rho_{AB}).
\end{equation}
To optimally distinguish between the channels, one must maximize the distinguishability between $ \rho_{AB}^1$ and $ \rho_{AB}^2$ over all possible input states $\rho_{AB}$. This optimization captures the advantage offered by entangled inputs in channel discrimination tasks. Hence, the distance between the two channels $\mathcal{E}_1$ and $\mathcal{E}_2$ (each occurring with equal a priori probability) when optimized over all input quantum states $\rho_{AB}$ can be written as

\begin{equation}\label{maximization}
\begin{split}
    & \mathbb{D} (\{\mathcal{E}_1,\mathcal{E}_2 \}) \\
    =& \max_{\rho_{AB}} || (\mathrm{id}_A \otimes \mathcal{E}_1)(\rho_{AB})-  (\mathrm{id}_A \otimes \mathcal{E}_2)(\rho_{AB})||_1
\end{split}
   \end{equation}
Therefore, the maximum probability of successfully distinguishing between the two quantum channels, optimized over all possible input states and final measurements, is given by
\begin{equation}
    p_{\text{success}}(\{ \mathcal{E}_1, \mathcal{E}_2 \}) = \frac{1}{2} \left[1 +  \mathbb{D}(\{ \mathcal{E}_1, \mathcal{E}_2 \})  \right],
\end{equation}
We now illustrate the operational advantages of non-absolutely separable states, as manifested in the task of quantum channel discrimination. It is worth mentioning that the proofs of the results presented below follow techniques similar to those employed in \cite{piani2009all,bae2019more}. Before proceeding further, we first establish the following lemma.\\

\begin{lemma} \label{lemma2}
A state $\sigma_{AB} \in  \mathcal{D}({{\mathbf{C}}^d \otimes  {\mathbf{C}}^d})$ does not belong to $\mathbb{AS}$ if and only if there exists a positive, trace-preserving map $\Lambda_{\text{TP}}$ such that 
\begin{equation}
   (  \mathrm{id}_A \otimes \Lambda_{\text{TP}}) (\mathcal{U} \, \sigma_{AB} \, \mathcal{U}^{\dagger}) <0
\end{equation}
for some global unitary operator $\mathcal{U} \in \mathbb{U}$.
\end{lemma}
\proof It is known that if a state $\sigma_{AB} \notin \mathbb{AS}$, then there exists a unitary operator $\mathcal{U}$ such that the transformed state $\mathcal{U} \, \sigma_{AB} \, \mathcal{U}^{\dagger}$ becomes entangled. Consequently, there always exists a positive linear map $\Lambda:\mathcal{M}_d \rightarrow \mathcal{M}_d $ for which
\begin{equation}
( \mathrm{id}_A \otimes \Lambda)\big( \mathcal{U} \, \sigma_{AB} \, \mathcal{U}^{\dagger} \big) < 0,
\end{equation}
Let $\mu(\Lambda) = \max_{\rho} \Tr[\Lambda(\rho)]$, where $\rho \in \mathcal{D}(\mathbf{C}^d)$. Define the normalized map $\Lambda^{\prime} = \Lambda / \mu(\Lambda)$. By construction, this ensures that
\begin{equation}
\Tr(\rho) \geq \Tr\big(\Lambda^{\prime}(\rho)\big) \quad \text{for all} \quad \rho \geq 0.
\end{equation}
Using this construction, we define a trace-preserving map $\Lambda_{\text{TP}}$ as
\begin{equation}
\Lambda_{\text{TP}}(\rho) = \Lambda^{\prime}(\rho) + \left[ \Tr(\rho) - \Tr\big(\Lambda^{\prime}(\rho)\big) \right] \ket{0}\bra{0},
\end{equation}
where the auxiliary state $\ket{0}$ is orthogonal to the output subspace of $\mathcal{M}_d$. Due to this orthogonality, the following equivalence holds:
\begin{equation}
\begin{split}
    &( \mathrm{id}_A \otimes \Lambda)\big( \mathcal{U} \, \sigma_{AB} \, \mathcal{U}^{\dagger} \big) < 0 \iff ( \mathrm{id}_A \otimes \Lambda_{\text{TP}})( \mathcal{U} \, \sigma_{AB} \, \mathcal{U}^{\dagger}) \\
    & \hspace{7cm} < 0
\end{split}
\end{equation}
for some global unitary operator $\mathcal{U} \in \mathbb{U}$.

This completes our proof. \qed

\begin{theorem} \label{theorem5}
A quantum state $\sigma_{AB} \notin \mathbb{AS}$ if and only if there exists two quantum channels $\mathcal{E}_1$ and $\mathcal{E}_2$ such that 
\begin{equation}
\begin{split}
     &||  (\mathrm{id}_A \otimes \mathcal{E}_1) (\mathcal{U} \, \sigma_{AB} \, \mathcal{U}^{\dagger}) -  (\mathrm{id}_A \otimes \mathcal{E}_2) (\mathcal{U} \, \sigma_{AB} \, \mathcal{U}^{\dagger}) ||_1 \\
    & \hspace{5cm}>\mathbb{D} (\{\mathcal{E}_1,\mathcal{E}_2 \})
\end{split}
    \end{equation}
    for some global unitary operator $\mathcal{U} \in \mathbb{U}$, where $\mathbb{D} (\{\mathcal{E}_1,\mathcal{E}_2 \})$ is defined in Eq. \eqref{maximization} with the maximization taken over all states $\rho_{AB} \in \mathbb{AS}$. 
\end{theorem}

\proof Let ${\Lambda}_{\text{TP}}:\mathcal{M}_d \rightarrow \mathcal{M}_d$ be a positive, trace-preserving, linear map. Now, from this trace-preserving map $({\Lambda}_{\text{TP}})$, one can construct a trace-annihilating map $({\Lambda}_{\text{TA}})$ as follows:
\begin{equation} \label{tpform}
    {\Lambda}_{\text{TA}} (\rho) = {\Lambda}_{\text{TP}} (\rho) - \text{Tr}(\rho) \ket{f}\bra{f}
\end{equation}
where, $\ket{f}\bra{f}$ is orthogonal to all elements of $\mathcal{M}_d$. From \cite{piani2009all}, we know that a trace-annihilating map $ {\Lambda}_{\text{TA}}$ can always be expressed as a scaled difference between two quantum channels $ \mathcal{E}_1$ and $ \mathcal{E}_2$, i.e. 
\begin{equation}\label{jordan}
    \mathcal{E}_1 - \mathcal{E}_2 = k \, {\Lambda}_{\text{TA}}
\end{equation}
where $k$ is a positive constant. Now, using Eq. \eqref{tpform} and Eq. \eqref{jordan}, we obtain
\begin{equation}
     \mathcal{E}_1 - \mathcal{E}_2 = k\, ({\Lambda}_{\text{TP}}  - \text{Tr}(.) \ket{f}\bra{f})
\end{equation}
Now consider a bipartite quantum state $\rho_{AB} \in \mathbb{AS}$ which implies that for any global unitary $\mathcal{U}$, the state $\mathcal{U} \, \rho_{AB} \, \mathcal{U}^{\dagger}$ remains separable, i.e., $\mathcal{U} \, \rho_{AB} \, \mathcal{U}^{\dagger} \in \mathbb{S}$. Then,
\begin{align} \label{abs}
    & || ( \mathrm{id}_A \otimes \mathcal{E}_1) (\mathcal{U} \, \rho_{AB} \, \mathcal{U}^{\dagger}) - ( \mathrm{id}_A \otimes \mathcal{E}_2) (\mathcal{U} \, \rho_{AB} \, \mathcal{U}^{\dagger}) ||_1 \\ \nonumber
     =& ||  (\mathrm{id}_A \otimes (\mathcal{E}_1 - \mathcal{E}_2)) (\mathcal{U} \, \rho_{AB} \, \mathcal{U}^{\dagger}) ||_1 \\ \nonumber
     =& k|| ( \mathrm{id}_A \otimes   {\Lambda}_{\text{TP}})(\mathcal{U} \, \rho_{AB} \, \mathcal{U}^{\dagger})  - \text{Tr}_{B} (\mathcal{U} \, \rho_{AB} \, \mathcal{U}^{\dagger}) \otimes \ket{f}\bra{f})  ||_1\\ \nonumber
     =&  k(|| ( \mathrm{id}_A \otimes   {\Lambda}_{\text{TP}})(\mathcal{U} \, \rho_{AB} \, \mathcal{U}^{\dagger})||_1 + 1)\\ \nonumber
     =&2k \nonumber
\end{align}
where the last step follows from the fact that $\Lambda_{\text{TP}}$ is trace-preserving, so the output has unit trace.\\

   Now consider a state $\sigma_{AB} \notin \mathbb{AS}$. By definition, there exists a global unitary $\mathcal{U} \in \mathbb{U}$ such that $\mathcal{U} \,\sigma_{AB} \, \mathcal{U}^\dagger$ is entangled. In this case, we have
    \begin{align} \label{nonabsadvantage}
    & ||  (\mathrm{id}_A \otimes \mathcal{E}_1) (\mathcal{U} \, \sigma_{AB} \, \mathcal{U}^{\dagger}) -  (\mathrm{id}_A \otimes \mathcal{E}_2) (\mathcal{U} \, \sigma_{AB} \, \mathcal{U}^{\dagger}) ||_1 \\ \nonumber
       =& k|| ( \mathrm{id}_A \otimes   {\Lambda}_{\text{TP}})(\mathcal{U} \, \sigma_{AB} \, \mathcal{U}^{\dagger})  - \text{Tr}_B(\mathcal{U} \, \sigma_{AB} \, \mathcal{U}^{\dagger}) \otimes \ket{f}\bra{f})  ||_1\\ \nonumber
     =&  k(|| ( \mathrm{id}_A \otimes   {\Lambda}_{\text{TP}})(\mathcal{U} \, \sigma_{AB} \, \mathcal{U}^{\dagger})||_1 + 1)\\ \nonumber
     >&2k \nonumber
\end{align}
since $\sigma_{AB}$ becomes entangled under the action of a global unitary operation $\mathcal{U}$ and hence the trace norm of $( \mathrm{id}_A \otimes \Lambda_{\text{TP}})(\mathcal{U} \, \sigma_{AB} \, \mathcal{U}^{\dagger})$ exceeds $1$, which proves the theorem. \qed\\

\begin{example}
    Consider the two-qubit state
\begin{equation}
\sigma_{AB}^1 = p \ket{{\phi}^{+}_1}\bra{{\phi}^{+}_1} + \frac{1 - p}{4} I_4,
\end{equation}
where $\ket{{\phi}^{+}_1} = \ket{00} \in \mathbb{C}^2 \otimes \mathbb{C}^2$ and $p \in [0,1]$.
\end{example}Since $\ket{{\phi}^{+}_1}$ is a product state and the remaining part is the maximally mixed state, therefore $\sigma_{AB}^1$ is separable for all $p \in [0,1]$. \

Now, consider the action of a global unitary operator defined by \begin{equation} \begin{split} \mathcal{U}_1 = \frac{1}{\sqrt{2}} \begin{pmatrix} 1 & 0 & 1 & 0 \\[4pt] 0 & 1 & 0 & 1 \\[4pt] 0 & 1 & 0 & -1 \\[4pt] 1 & 0 & -1 & 0 \end{pmatrix}. \end{split} \end{equation} 

Applying this global unitary to $\sigma_{AB}^1$ yields the state \begin{equation} \sigma_{AB}^2 = \mathcal{U}_1 \sigma_{AB}^1 \mathcal{U}_1^\dagger = p \ket{{\phi}^{+}_2} \bra{{\phi}^{+}_2} + \frac{1 - p}{4} I_4, \end{equation} where $\ket{{\phi}^{+}_2} = \frac{1}{\sqrt{2}}(\ket{00} + \ket{11}) \in \mathbb{C}^2 \otimes \mathbb{C}^2$. Using the partial transposition criterion, we find that $(\sigma_{AB}^2)^{T_A}$ possesses a negative eigenvalue whenever $p > \frac{1}{3}$. This implies that, $\sigma_{AB}^2 $ is entangled for $p > \frac{1}{3}$. Moreover, this observation guarantees that $\sigma_{AB}^1$ is a separable but not absolutely separable state. Therefore similarly from Eq.\eqref{nonabsadvantage}, we get
 \begin{align} 
 & || (\mathrm{id}_A \otimes \mathcal{E}_1) (\mathcal{U} \,  \sigma_{AB}^1 \, \mathcal{U}^{\dagger}) - (\mathrm{id}_A \otimes \mathcal{E}_2) (\mathcal{U} \,  \sigma_{AB}^1 \, \mathcal{U}^{\dagger}) ||_1 \\ \nonumber 
 =& k|| ( \mathrm{id}_A \otimes {\Lambda}_{\text{TP}})(\mathcal{U} \,  \sigma_{AB}^1 \, \mathcal{U}^{\dagger}) - \text{Tr}_B(\mathcal{U} \,  \sigma_{AB}^1 \, \mathcal{U}^{\dagger}) \otimes \ket{f}\bra{f}) ||_1\\
 \nonumber 
 =& k(|| ( \mathrm{id}_A \otimes {\Lambda}_{\text{TP}})(\mathcal{U} \,  \sigma_{AB}^1 \, \mathcal{U}^{\dagger})||_1 + 1)\\ \nonumber
 >&2k  
 \end{align} 
 since $\sigma_{AB}^1$ is separable but not absolutely separable, hence the trace norm of $( \mathrm{id}_A \otimes \Lambda_{\text{TP}})(\mathcal{U} \,  \sigma_{AB}^1 \, \mathcal{U}^{\dagger})$ exceeds $1$.

 However, for arbitrary bipartite quantum state $\rho_{AB} \in \mathbb{AS}$ and for any global unitary $\mathcal{U} \in \mathbb{U}$, from Eq. \eqref{abs} we obtain \begin{align} 
  || ( \mathrm{id}_A \otimes \mathcal{E}_1) (\mathcal{U} \, \rho_{AB} \, \mathcal{U}^{\dagger}) - ( \mathrm{id}_A \otimes \mathcal{E}_2) (\mathcal{U} \, \rho_{AB} \, \mathcal{U}^{\dagger}) ||_1  =2k \nonumber 
 \end{align} 
  This example demonstrates explicitly that separable but not absolutely separable states
can provide a strictly greater advantage in bipartite channel discrimination compared to any absolutely separable state, thereby confirming the operational significance highlighted in Theorem 7.

\section{Conclusions}\label{s6} 
Non-absolutely separable states represent a useful resource in various quantum information processing tasks, due to their potential of transforming themselves into entangled states via suitable global unitary operations. However, before such states can be harnessed for practical use, it is of foremost importance to reliably detect their presence. In this work, we introduce a moment-based detection framework to identify the signatures of non-absolute separability. Our method is based on the evaluation of simple, efficiently computable functionals that do not require full spectral information of the unknown quantum state, thereby avoiding the need for full state tomography \cite{aaronson2018shadow,aaronson2019gentle,huang2020predicting,elben2020mixed}. This highlights a significant advantage of our criteria in comparison to the eigenvalue spectrum-based criterion of \cite{PhysRevA.88.062330}, where full state tomography is necessary. Notably, unlike traditional witness-based approaches \cite{PhysRevA.89.052304}, our protocol remains effective even when only partial information about the state is available. We further support our approach by providing concrete examples that demonstrate the effectiveness of our proposed detection scheme. 

A related challenge lies in identifying states that retain the positive partial transpose property under all non-local unitary transformations—referred to as absolutely PPT states. A necessary and sufficient condition for detecting such states, based on eigenvalues, was established in \cite{PhysRevA.76.052325}. However, the application of this criterion becomes increasingly challenging as the system dimension grows. In this work, we propose a moment-based framework for the detection of non-absolutely PPT states, providing a scalable and conceptually distinct alternative to existing approaches.

Subsequently, we explore our proposed moment-based criteria for detecting non-absolutely separating channels, as these channels enable the recovery of entanglement via suitable global unitary transformations. This, in turn, facilitates the identification of channels that retain operational utility across a broad class of quantum information protocols. Lastly, we examine the operational advantage of non-absolutely separable states in the context of quantum channel discrimination tasks. We show that every non-absolutely separable state offers an advantage over all absolutely separable states to discriminate at least one pair of quantum channels, i.e., for each such state, there exists a specific discrimination scenario in which it yields a strictly higher success probability than any absolutely separable state. 

It is worth emphasizing that certifying absolute properties is a fundamentally hard problem, as it requires verification over the entire global unitary orbit of a state; this complexity is intrinsic to the notion of absolute separability and cannot be fully avoided. Our criteria provide sufficient conditions for identifying states that are not absolutely separable (or not absolutely PPT) by certifying the existence of at least one global unitary under which the state becomes entangled (or NPT). Note that these criteria do not reduce the worst-case computational complexity of exploring the full unitary orbit, nor do they single out a unique optimal unitary. Instead, they recast the detection of non-absolute properties in terms of moment-based conditions, which are physically motivated and experimentally accessible. Although the fundamental complexity remains unchanged, the use of moments enables practical detection schemes based on modern techniques such as randomized measurements and shadow tomography techniques \cite{aaronson2018shadow,aaronson2019gentle,huang2020predicting,elben2020mixed}. 

This work opens several compelling avenues for future investigation. Since our method provides a necessary condition for absolute separability, a key challenge moving forward is to establish necessary and sufficient conditions for detecting non-absolute separability in an efficient way. Furthermore, the versatility of our proposed moment-based framework suggests that it may be extended to identify other forms of non-absoluteness in quantum states \cite{patro2017non,patro2022absolute,patro2024quantum}, which we leave as a direction for future investigation. Given the practical implementability of our detection protocol, its experimental realization emerges as a natural outcome of our present analysis.

\section{Acknowledgements}
B.M. acknowledges the DST INSPIRE fellowship program for financial support. N.G. acknowledges support 
from DST-ANRF (SERB) MATRICS Grant No. MTR/2022/000101.
\bibliography{main}

\end{document}